\documentclass{webofc}

\usepackage[varg]{txfonts}   
\usepackage{hyperref}
\usepackage{url}

\hypersetup{colorlinks=true,citecolor=blue,urlcolor=blue,linkcolor=blue}
\begin{document}
\title{Heavy flavor angular correlations as probes of the glasma}
%
%

\author{
\firstname{Dana} \lastname{Avramescu}\inst{1,2}\fnsep\thanks{\email{dana.d.avramescu@jyu.fi}} 
\and
\firstname{Vincenzo} \lastname{Greco}\inst{3,4}\fnsep\thanks{\email{greco@lns.infn.it}} 
\and
\firstname{Tuomas} \lastname{Lappi}\inst{1,2}\fnsep\thanks{\email{tuomas.v.v.lappi@jyu.fi}}
\and
\firstname{Heikki} \lastname{M\"{a}ntysaari}\inst{1,2}\fnsep\thanks{\email{heikki.mantysaari@jyu.fi}}
\and
\firstname{David} \lastname{M\"{u}ller}\inst{5}\fnsep\thanks{\email{dmueller@hep.itp.tuwien.ac.at}}
}

\institute{
Department of Physics, University of Jyväskylä,  P.O. Box 35, 40014 University of Jyväskylä, Finland
\and
Helsinki Institute of Physics, P.O. Box 64, 00014 University of Helsinki, Finland 
\and
Department of Physics and Astronomy, University of Catania, Via S. Sofia 64, I-95123 Catania, Italy
\and
INFN-Laboratori Nazionali del Sud, Via S. Sofia 62, I-95123 Catania, Italy
\and 
Institute for Theoretical Physics, TU Wien, Wiedner Hauptstraße 8, A-1040 Vienna, Austria
}

\abstract{We study the effect of the glasma fields, formed in the early stage of heavy-ion collisions, on the transport of $Q\bar{Q}$ pairs produced back-to-back. We find that for pairs with moderate initial transverse momentum $p_T$ evolving in glasma fields with sufficiently large saturation momentum $Q_s$, the azimuthal correlation $\mathcal{C}({\Delta\phi})$ is quickly affected. The decorrelation widths $\sigma_{\Delta\phi}$ during the glasma and Quark Gluon Plasma (QGP) phases are comparable. 
}
\maketitle
\section{Introduction}
\label{intro}
At weak coupling, the early stage of heavy-ion collisions is described using the Color Glass Condensate (CGC) framework. In this approach the collision of saturated gluon fields generates a non-equilibrium glasma state.
As heavy quarks are produced early, they can provide a sensitive probe of the glasma. The classical transport of heavy quarks in glasma fields has been used to investigate the effect of the glasma on the nuclear modification factor in pA \cite{Ruggieri:2018rzi} and AA \cite{Sun:2019fud,Gale:2024cdz} collisions, and $Q\bar{Q}$ pairs dissociation in glasma \cite{Pooja:2024rnn,Oliva:2024rex}. In this work, we focus on how the early glasma fields deflect $Q\bar{Q}$ pairs and alter their azimuthal correlations \cite{Avramescu:2024xts}. 

\section{Heavy quark pairs in glasma}
\label{sec-1}
The glasma fields are obtained by numerically solving the sourceless classical Yang-Mills (CYM) equations for the gluon fields $A^\mu$, namely $\mathcal{D}_\mu F^{\mu\nu}=0$, with the covariant derivative $\mathcal{D}_\mu=\partial_\mu-\mathrm{i}gA_\mu$ and the field strength tensor $F^{\mu\nu}=\partial^\mu A^\nu-\partial^\nu A^\mu-\mathrm{i}g[A^\mu, A^\nu]$. The initial condition for the glasma fields is given in terms of CGC fields. The fields are sampled from classical color charge sources according to the MV model $\langle \rho^a(\boldsymbol{x}_T)\,\rho^b(\boldsymbol{y}_T)\rangle = (g^2\mu)^2\delta^{ab}\delta^{(2)}(\boldsymbol{x}_T-\boldsymbol{y}_T)$, with $g^2\mu\propto Q_s$ proportional to the saturation momentum $Q_s$. The field equations of motion are solved using a non-Abelian real-time lattice discretization. 

\begin{figure*}
\centering
\includegraphics[width=\textwidth,clip]{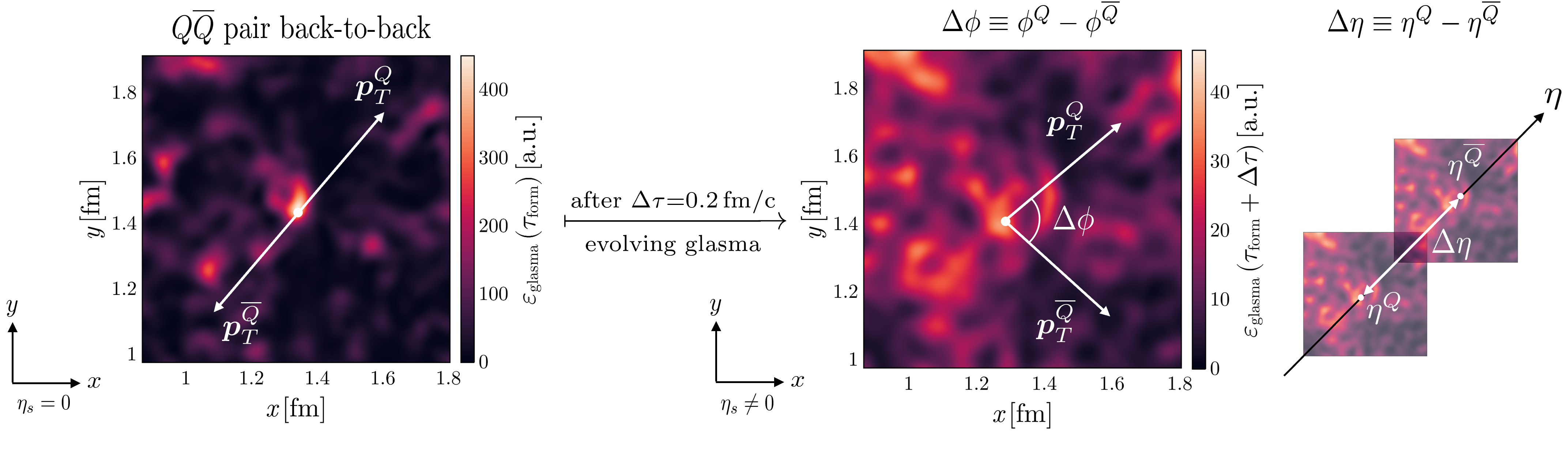}
\caption{Sketch depicting a $Q\bar{Q}$ pair with momenta $\boldsymbol{p}_T^Q=-\boldsymbol{p}_T^{\bar{Q}}$ at mid-rapidity $\eta=0$ (\textit{left}). Due to the glasma fields (\textit{background}) the quarks in the pair experience a change in $\Delta\phi$ (\textit{middle}) and $\Delta\eta$ (\textit{right}).}
\label{fig:qqbar}       
\end{figure*}

Heavy quarks obey Wong's classical transport equations for test particles in a CYM background gauge field $A^\mu$, in this case the glasma, expressed as 
\begin{equation}
    \label{eq:wong}
    \frac{\mathrm{d}x^\mu}{\mathrm{d}\tau}=\frac{p^\mu}{m},\quad \dfrac{\mathrm{d}p^\mu}{\mathrm{d}\tau}=\frac{g}{T_R}\mathrm{Tr}\{QF^{\mu\nu}\}\frac{p_\nu}{m},\quad \frac{\mathrm{d}Q}{\mathrm{d}\tau}=-\mathrm{i}g[A_\mu,Q]\frac{p^\mu}{m},
\end{equation}
where $x^\mu$ is the coordinate, $p^\mu$ the momentum and $Q=Q^a T^a$ the classical color charge for a heavy quark. Here $\mathrm{Tr}\{T^aT^b\}=T_R\delta^{ab}$ with $T_F=1/2$ for $R=F$ quarks in the fundamental representation of SU($3$). Both the particle and field equations of motion are solved numerically \cite{Avramescu:2023qvv}. We simulate quark $Q$ and anti-quark $\bar{Q}$ pairs in the glasma according to Wong's equations from Eq. \eqref{eq:wong}. The pairs are initialized at the same coordinate $\boldsymbol{x}_T(Q)=\boldsymbol{x}_T({\bar{Q}})$ and $\eta({Q,\bar{Q}})=0$, with opposite transverse momenta $\boldsymbol{p}_T(Q)=-\boldsymbol{p}_T({\bar{Q}})$ (leading order perturbative QCD production), null longitudinal momentum $p^\eta({Q,\bar{Q}})=0$, and random color charge vectors (the heavy quark pairs are mostly produced from gluon fusion $gg\rightarrow Q\bar{Q}$). Each pair is formed at $\tau_\mathrm{form}= 1/(2m_T)$ with the transverse mass $m_T^2=p_T^2+m^2_\mathrm{HQ}$ and initialized with a finite $\boldsymbol{p}_T(\tau_\mathrm{form})$. The $Q\bar{Q}$ pairs evolving in the glasma are illustrated in Fig.~\ref{fig:qqbar}. 

\section{Heavy flavor angular correlations}
\label{sec-2}

\begin{figure*}[t]
\centering
\includegraphics[width=0.3\columnwidth]
{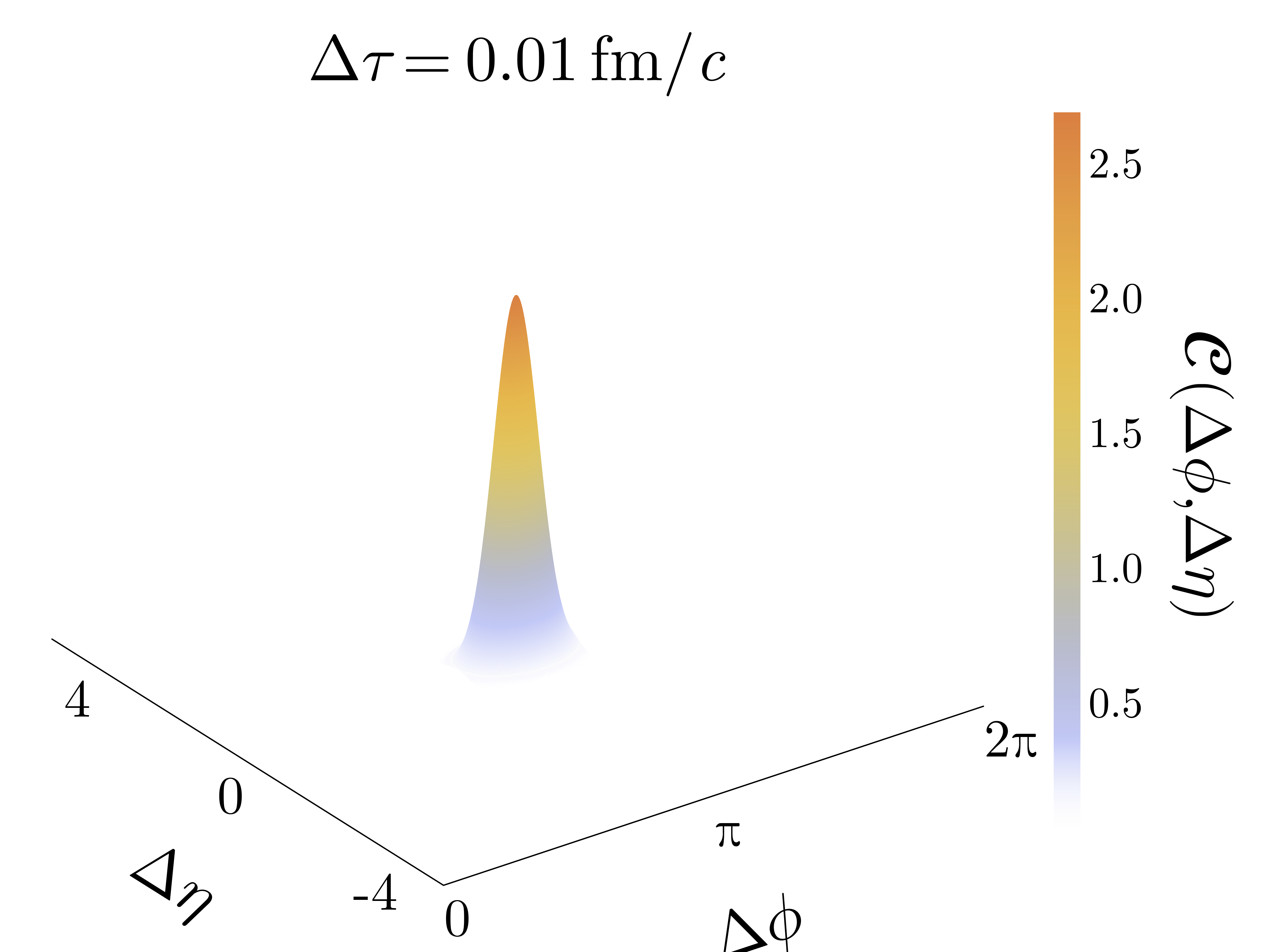}\quad
\includegraphics[width=0.3\columnwidth]
{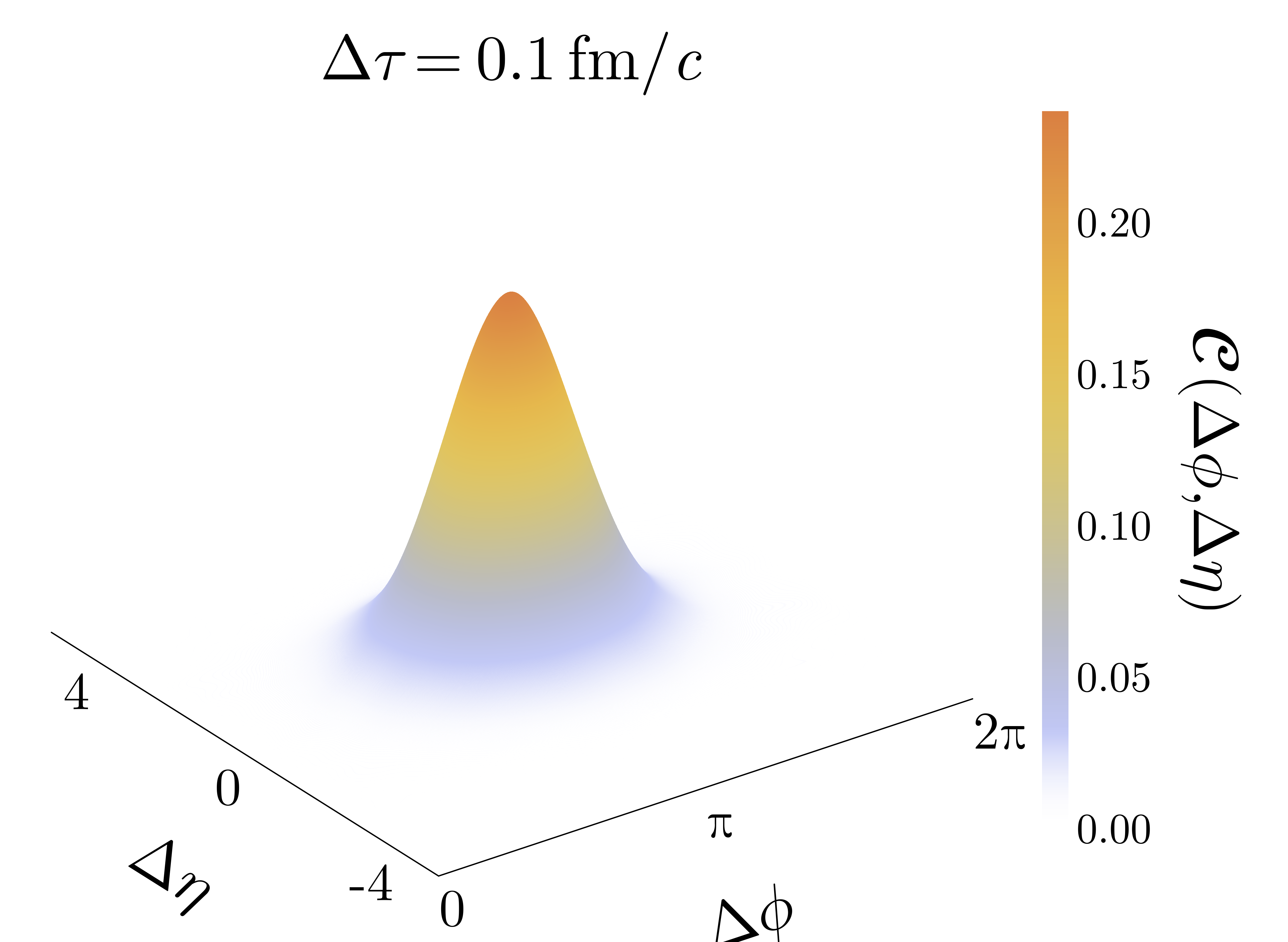}\quad
\includegraphics[width=0.3\columnwidth]
{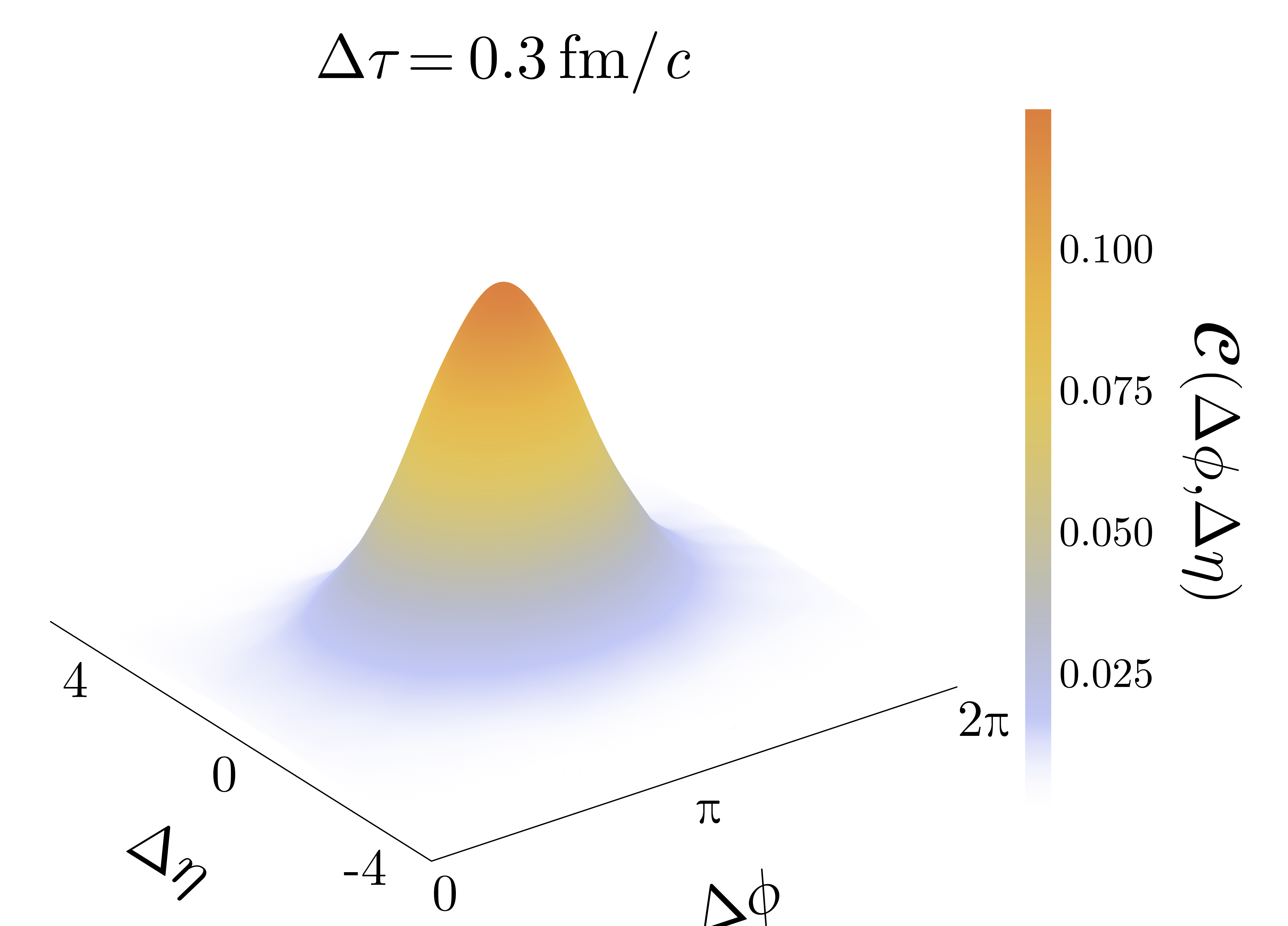}
\caption{Snapshots at various $\Delta\tau$ of the $c\bar c$ correlation in relative rapidity $\Delta \eta$ and azimuthal angle $\Delta\phi$.}
\label{fig:3dtwopartcorr}
\end{figure*}

\begin{figure*}
\centering
\includegraphics[width=0.7\textwidth,clip]{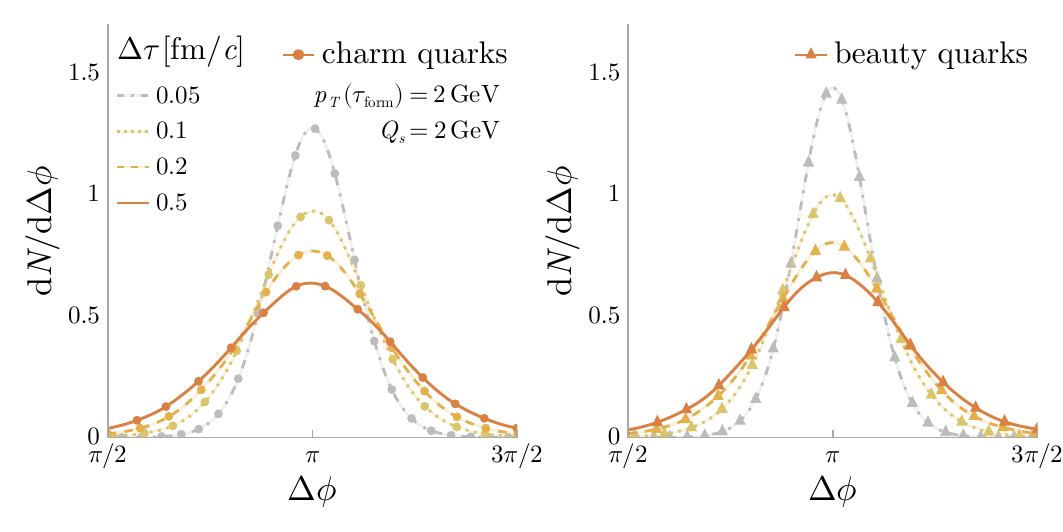}
\caption{Azimuthal correlation $\mathcal{C}(\Delta\phi)$ for $c\bar{c}$ (\textit{left}) and $b\bar{b}$ (\textit{right}) pairs with initial $p_T(\tau_\mathrm{form})=2\,\mathrm{GeV}$ at $\Delta\tau\in\{0.05, 0.1, 0.2, 0.5\}\,\mathrm{fm/c}$ (\textit{color and line style}), in a glasma with $Q_s=2\,\mathrm{GeV}$.}
\label{fig:dndphi_tau}      
\end{figure*}

\begin{figure*}
\centering
\vspace*{1cm}      
\includegraphics[width=0.95\textwidth,clip]{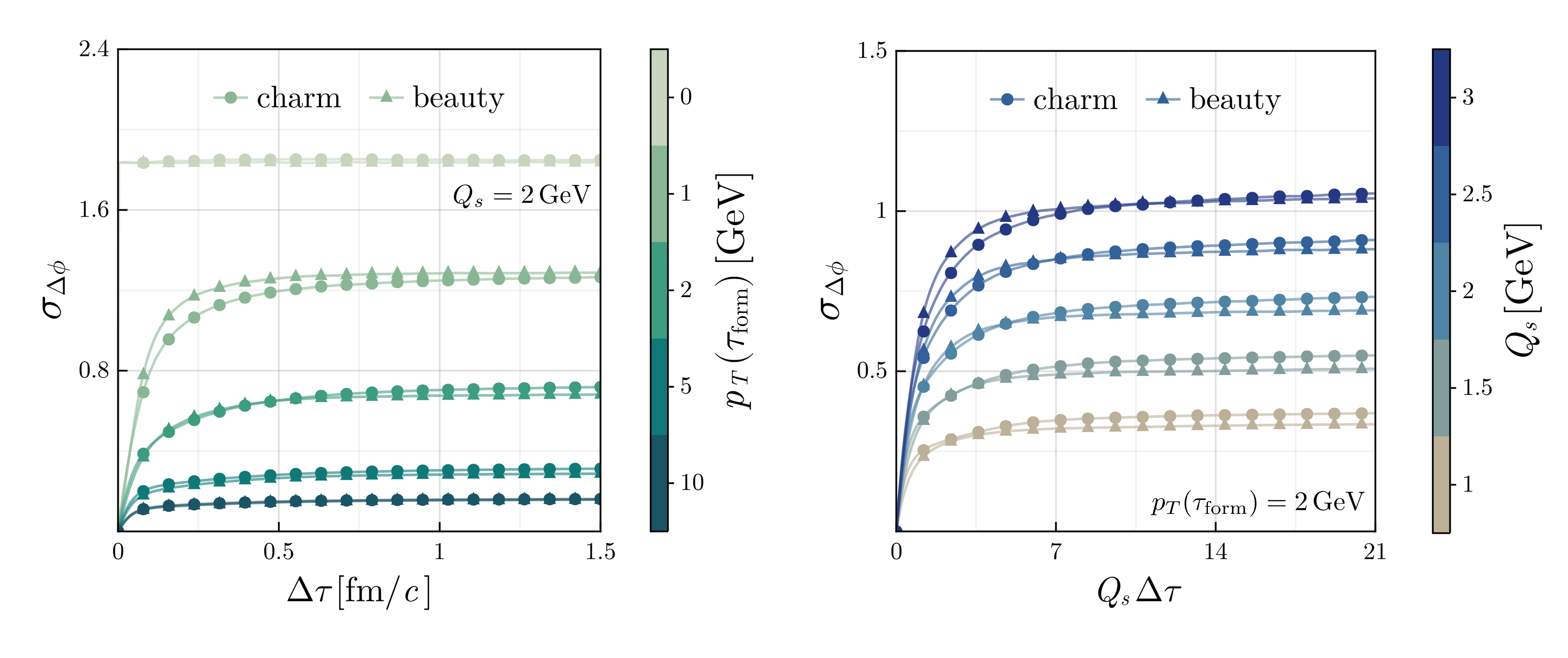}
\caption{Correlation width in azimuthal angle $\sigma_{\Delta\phi}$ for $c\bar{c}$ (\textit{circle markers}) and $b\bar{b}$ (\textit{triangle markers}) pairs as a function of $\Delta\tau$ for different $p_T$ (\textit{left}) and $Q_s$ (\textit{right}) values.}
\label{fig:sigma_dphi_pT_Qs_dep}       
\end{figure*}

After formation, the pairs instantaneously interact with the glasma fields and experience momentum deflections due to the color Lorentz force. This leads to a change in relative pseudorapidity $\Delta\eta= \eta(Q)-\eta({\bar{Q}})$ where $\eta({q})=\ln\{(p^q+p_z^q)/(p^q-p_z^q)\}/2$ with $(p^q)^2=(p_T^q)^2+(p_z^q)^2$ for $q\in\{Q,\bar{Q}\}$, and the relative  azimuthal angle $\Delta\phi$ between $\boldsymbol{p}_T(Q)$ and $\boldsymbol{p}_T(\bar{Q})$. 
We simulate many test particles in multiple glasma events and extract the two-particle correlation
\begin{equation}
    \label{eq:twopartcorr}
    \mathcal{C}(\Delta\eta,\Delta\phi)= \frac{1}{N_\mathrm{pairs}}\frac{\mathrm{d}^2N}{\mathrm{d}\Delta\eta\,\mathrm{d}\Delta\phi},
\end{equation} 
as a function of the relative proper time $\Delta\tau= \tau-\tau_\mathrm{form}$. Results from numerical simulations are shown in Fig.~\ref{fig:3dtwopartcorr} for $c\bar{c}$ pairs at different $\Delta\tau$ values. The magnitude of the correlation $\mathcal{C}(\Delta\phi,\Delta\eta)$ rapidly drops after $\Delta\tau=1/Q_s\approx 0.1\,\mathrm{fm/c}$ for $Q_s=2\,\mathrm{GeV}$. By integrating Eq.~(\ref{eq:twopartcorr}) along $\Delta\eta$ we extract the azimuthal correlations $\mathcal{C}(\Delta\eta)= 1/N_\mathrm{pairs}\,\mathrm{d}N/\mathrm{d}\Delta\eta$ as a function of $\Delta\tau$. Results for $c\bar{c}$ and $b\bar{b}$ pairs are shown in Fig.~\ref{fig:dndphi_tau}. The initial correlation at $\Delta\phi=\pi$ quickly decreases after $\Delta\tau=0.05\,\mathrm{fm/c}$ and the decrease gets slower with increasing time. The decorrelation is similar for charm and beauty quarks, showing that this is a glasma effect not influenced by the particle mass or formation time. Further, we quantify the magnitude of the correlation through the width  $\sigma_{\Delta\phi}$ computed as the standard deviation of the distribution $\mathcal{C}(\Delta\phi)$. Fig.~\ref{fig:sigma_dphi_pT_Qs_dep} depicts $\sigma_{\Delta\phi}$ for both charm and beauty quark pairs, as a function of $\Delta\tau$. Pairs with small initial $p_T$ immediately get decorreated by the glasma fields. The decorrelation is more pronounced for glasma fields characterized by larger $Q_s$. These observations hold for both $c\bar{c}$ and $b\bar{b}$ pairs. The decorrelation width at $Q_s=2\,\mathrm{GeV}$ for $c\bar{c}$ pairs with $p_T\in[5,10]\,\mathrm{GeV}$ at $\Delta\tau=0.3\,\mathrm{fm/c}$ is $\sigma_{\Delta\phi}^\mathrm{glasma}\approx 0.3$. The value reported during the QGP for $c\bar{c}$ pairs with $p_T\in[4,10]\,\mathrm{GeV}$ (value with collisional and radiative energy loss included in the heavy quark transport model)  yields $\sigma_{\Delta\phi}^\mathrm{QGP}\approx 0.36$~\cite{Nahrgang:2013saa}, roughly the same order of magnitude as the glasma effect.

\label{summ}
\section{Conclusion}
We studied the classical transport of $c\bar{c}$ (and $b\bar{b}$) pairs in the glasma early-stage fields. The initially strong glasma color fields induce a large decorrelation in the azimuthal angle, with the decorrelation width $\sigma_{
\Delta\phi}$ in glasma comparable to the value accumulated during the next stages. The study of $c\bar{c}$ angular correlations, and disentangling between initial state effects and the QGP contribution,  will be relevant in making theoretical predictions for future $D\bar{D}$ correlations, planned to be measured during ALICE3 \cite{ALICE:2022wwr}. 

\noindent\textit{\textbf{Acknowledgments.}} This work was supported by the Research Council of Finland, the Centre of Excellence in Quark Matter (projects 346324 and 364191), and projects 338263, 346567, and 359902, and by the European Research Council (ERC, grant agreements ERC-2023-101123801 GlueSatLight and ERC-2018-ADG-835105 YoctoLHC). D.A. acknowledges the support of the Vilho, Yrj\"{o} and Kalle V\"{a}is\"{a}l\"{a} Foundation. D.M. acknowledges support from the Austrian Science Fund (FWF) projects P 34764 and P 34455.


\begin{thebibliography}{}
\bibitem{Ruggieri:2018rzi}
M. Ruggieri, S. K. Das, Cathode tube effect: Heavy quarks probing the glasma in p-Pb collisions. \href{https://doi.org/10.1103/PhysRevD.98.094024}{Phys. Rev. D \textbf{98}, 094024 (2018)}

\bibitem{Sun:2019fud}
Y. Sun, G. Coci, S. K. Das, S. Plumari, M. Ruggieri, V. Greco, Impact of Glasma on heavy quark observables in nucleus-nucleus collisions at LHC. \href{https://doi.org/10.1016/j.physletb.2019.134933}{Phys. Lett. B \textbf{798}, 134933 (2019)}

\bibitem{Gale:2024cdz}
C. Gale, S. Jeon, M. Kurian, B. Schenke, M. Singh, Charm quark evolution in the early stages of heavy-ion collisions. \href{https://arxiv.org/abs/2509.18647}{arXiv:2509.18647}.

\bibitem{Pooja:2024rnn}
Pooja, M. J. Jamal, P. P. Bhaduri, M. Ruggieri, S. K. Das, $c\bar{c}$ and $b\bar{b}$ suppression in the glasma. \href{https://doi.org/10.1103/PhysRevD.110.094018}{Phys. Rev. D \textbf{110}, 094018 (2024)}.

\bibitem{Oliva:2024rex}
L. Oliva, G. Parisi, M. Ruggieri, Melting of $c\bar{c}$ and $b\bar{b}$ pairs in the pre-equilibrium stage of proton-nucleus collisions at the Large Hadron Collider. \href{https://doi.org/10.1103/nc3z-vns9}{Phys. Rev. D \textbf{112}, 014008 (2025)}.

\bibitem{Avramescu:2024xts}
D. Avramescu, V. Greco, T. Lappi, H. Mäntysaari, D. Müller, Heavy flavor angular correlations as a direct probe of the glasma. \href{https://doi.org/10.1103/PhysRevLett.134.172301}{Phys. Rev. Lett. \textbf{134}, 172301 (2025)}; The impact of glasma on heavy flavor azimuthal correlations and spectra. \href{https://doi.org/10.1103/PhysRevD.111.074036}{Phys. Rev. D \textbf{111}, 074036 (2025)}.

\bibitem{Avramescu:2023qvv}
D. Avramescu,  V. Băran, V. Greco, A. Ipp, D. Müller, M. Ruggieri, Simulating jets and heavy quarks in the glasma using the colored particle-in-cell method. \href{https://doi.org/10.1103/PhysRevD.107.114021}{Phys. Rev. D \textbf{107}, 114021 (2023)}.

\bibitem{Nahrgang:2013saa}
M. Nahrgang, J. Aichelin, P. B. Gossiaux, K. Werner, Azimuthal correlations of heavy quarks in Pb + Pb collisions at $\sqrt{s}=2.76$ TeV at the CERN Large Hadron Collider. \href{https://doi.org/10.1103/PhysRevC.90.024907}{Phys. Rev. C \textbf{90}, 024907 (2020)}. 

\bibitem{ALICE:2022wwr}
ALICE Collaboration, Letter of intent for ALICE 3: A next-generation heavy-ion experiment at the LHC. \href{https://doi.org/10.48550/arXiv.2211.02491}{arXiv:2211.02491}
\end{thebibliography}
\end{document}